\documentclass[twocolumn, superscriptaddress, prl, showpacs]{revtex4}

\usepackage{graphicx}
\usepackage{amsmath}

\def\beq{\begin{equation}}
\def\eeq{\end{equation}}
\def\bea{\begin{eqnarray}}
\def\eea{\end{eqnarray}}
\def\nn{\nonumber\\}
\def\r{{\bf r}}
\def\rb{{\bar{\bf r}}}
\def\v{{\bf v}}
\def\k{{\bf k}}
\def\R{{\bf R}}
\def\I{{\bf I}}
\def\A{{\bf A}}
\def\M{{\bf M}}
\def\m{{\bf m}}
\def\0{{\bf 0}}
\def\J{{\bf J}}
\def\P{{\bf P}}
\def\bij{{\ev{ij}}}
\def\bji{{\ev{ji}}}
\def\sLC{_{\rm LC}}
\def\sIC{_{\rm IC}}
\def\nhat{\hat{\bf n}}

\def\sab{_{\alpha\beta}}

\def\parb{\partial_{k_\beta}}
\def\park{\partial_{\bf k}}
\def\ddkk{\frac{d^2k}{(2\pi)^2}}
\def\ltp{{\frac{A_0}{(2\pi)^2}}}
\def\ket#1{\vert#1\rangle}

\def\me#1#2#3{\langle#1\vert#2\vert#3\rangle}
\def\ev#1{\langle#1\rangle}

\begin{document}

\title{Orbital magnetization in periodic insulators}

\author{T. Thonhauser}
\affiliation{Department of Physics and Astronomy, Rutgers University, 
Piscataway, New Jersey 08854, USA}
\author{Davide Ceresoli}
\affiliation{International School for Advanced Studies (SISSA/ISAS)
and INFM-DEMOCRITOS, Trieste, Italy}
\author{David Vanderbilt}
\affiliation{Department of Physics and Astronomy, Rutgers University, 
Piscataway, New Jersey 08854, USA}
\author{R. Resta}
\affiliation{Dipartimento di Fisica Teorica Universit\`a di Trieste and
INFM-DEMOCRITOS, Trieste, Italy} 
\date{\today}

\begin{abstract}  
Working in the Wannier representation, we derive an expression for the
orbital magnetization of a periodic insulator.  The magnetization is
shown to be comprised of two contributions, an obvious one associated
with the internal circulation of bulk-like Wannier functions in the
interior, and an unexpected one arising from net currents carried by
Wannier functions near the surface.  Each contribution can be expressed
as a bulk property in terms of Bloch functions in a gauge-invariant
way.  Our expression is verified by comparing numerical tight-binding
calculations for finite and periodic samples.
\end{abstract}

\maketitle

Recent years have seen a surge of interest in issues of charge and spin
transport in magnetic materials and nanostructures, notably the
development of a theory of the intrinsic anomalous Hall conductivity
and some controversies surrounding the spin-Hall
effect~\cite{PhysicsToday}. In this context it is quite surprising that
the theory of orbital magnetization, essential for any proper
description of magnetism, has remained in a primitive state. 
Linear-response methods allow calculations of magnetization
\emph{changes}~\cite{linear, Sebastiani01, Mauri, Sebastiani02}, but
not of the magnetization itself.

Hirst~\cite{Hirst97} has emphasized that a knowledge of the bulk local
current density $\J(\r)$ is insufficient, even in principle, to
determine the macroscopic orbital magnetization $\M$, just as
the density $\rho(\r)$ cannot be used to
determine the electric polarization $\P$. 
Thus, the theory of $\M$ today is in a condition very similar to that
of $\P$ in the early 1990s, when the problem of computing \emph{finite}
polarization changes was solved by the introduction of the Berry-phase
theory~\cite{KSV,rap-a12}.  The essential difficulty, that the matrix
elements of the position operator $\r$ are not well-defined in the
Bloch representation, could be overcome by reformulating the problem in
the Wannier representation. Because Wannier functions (WFs) are
exponentially localized in an insulator, matrix elements of $\r$
between WFs are indeed well-defined.

Here we show that it is possible to formulate a corresponding theory of
the orbital magnetization for an insulator with broken time-reversal
symmetry.  The problem is analogous, with the circulation operator
${\r}\times{\v}$ now being ill-defined in the Bloch representation. 
Working instead in the Wannier representation, we write the orbital
magnetization as a gauge-invariant Brillouin-zone integral over
occupied Bloch functions. It contains two terms, the first of which
describes the internal circulation of bulk-like
WFs~\cite{ChemPhysChem}. The second is much more subtle, arising only
from surface WFs and reflecting the fact that the information about
surface currents needed to define the macroscopic magnetization is
actually contained in the bulk bandstructure. The resulting formula is
consistent with a recent semiclassical argument~\cite{Niu05} and can
easily be implemented in existing first-principles codes.

For our derivation, we restrict ourselves to the case of an insulator
described by a one-particle Hamiltonian with broken time-reversal
symmetry.  While the restriction to insulators is essential for the
theory of polarization, we suspect that it is less so here, so that
future generalizations to metals are not ruled out.  We also require a
vanishing macroscopic magnetic field (or, more generally, an integer
number of flux quanta per unit cell) so that the Bloch wavevector $\k$
remains a good quantum number. We have in mind cases in which a
staggered magnetic field averages to zero over the unit cell, or in
which the time-reversal breaking comes about through spin-orbit
coupling to a background of ordered local moments~\cite{Haldane88,
Ohgushi00, Jungwirth02, Murakami03, Yao04}.  For simplicity we work
with spinless electrons (the generalization to the spin-unrestricted
case being straightforward) and furthermore restrict ourselves to
zero-Chern-number insulators~\cite{Haldane88, Ohgushi00}.

Let us consider a finite sample representing a
fragment of a larger crystalline system. We assume that
the occupied states can be  represented in terms of well-localized
orthonormal orbitals $\ket{\phi_i}$, which we will refer to as Wannier
functions. If we introduce the velocity operator as
\beq
\v = -\frac{i}{\hbar}[\r,H]\;,
\label{vdef}
\eeq
then the total magnetic moment of the finite system involves the matrix
elements $\me{\psi_i}{\r\times\v}{\psi_i}$, where the $\ket{\psi_i}$
are the occupied eigenstates of $H$. By invariance of the trace, this
can be written in terms of WFs as
\beq
\m = -\frac{e}{2c}\sum_i \me{\phi_i}{\r\times\v}{\phi_i}\;,\label{m_tot_1}
\eeq
where $-e$ is the electron charge. The magnetization \M\ can then be
defined as the magnetic moment \m\ per unit volume. For large but
finite samples, all $\ket{\phi_i}$ that are sufficiently far from the
surface become exponentially similar to bulk WFs, which we will denote
as $\ket{w_i}$. For the electric polarization, the transition 
$\me{\phi_i}{\r}{\phi_i}\rightarrow\me{w_i}{\r}{w_i}$  describes the
polarization of periodic systems correctly~\cite{KSV}. Thus, it is
tempting to assume that the magnetization should be expressible in a
similar way in terms of the circulation $\me{w_i}{\r\times\v}{w_i}$ of
bulk WFs.

\begin{figure}
\begin{center}
\includegraphics[width=3.0in]{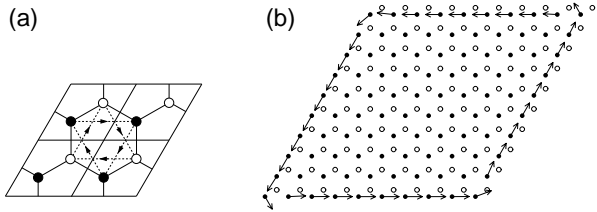}
\end{center}
\caption{\label{fig:model}{\bf (a)} Four unit cells of the Haldane
model.  Filled (open) circles denote sites with $E_0=-2$ ($+2$). Arrows
indicate sign of the phase $\varphi$ for second-neighbor hopping. {\bf
(b)} Net currents $-e\v_i$ associated with WFs, plotted at their
centers $\rb_i$, for a 10$\times$10 sample with $\varphi=\pi/4$.
Currents decrease rapidly into the bulk, so that only surface currents
are visible.}
\end{figure}

We set out to verify this hypothesis in the context of numerical
tight-binding calculations. For simplicity, we chose the Haldane 
model~\cite{Haldane88}, which is comprised of a honeycomb lattice with
two tight-binding sites per cell with site energies $\pm E_0$, real
first-neighbor hoppings $t_1$, and complex second-neighbor hoppings
$t_2e^{\pm i\varphi}$ as shown in Fig.~\ref{fig:model}(a). For our
tests, we have chosen a lattice constant equal to unity, $ E_0=\pm 2$,
$t_1=1$  and $t_2=1/3$ and allowed $\varphi$ to vary
\cite{para-choice}.  We treat the upper band as empty and the lower
band as occupied. The corresponding WFs were obtained by acting with
the band projector on a set of $\delta$-functions located on the sites
with $E_0=-2$ and applying a subsequent symmetric orthonormalization.

\begin{figure}
\includegraphics[width=2.8in]{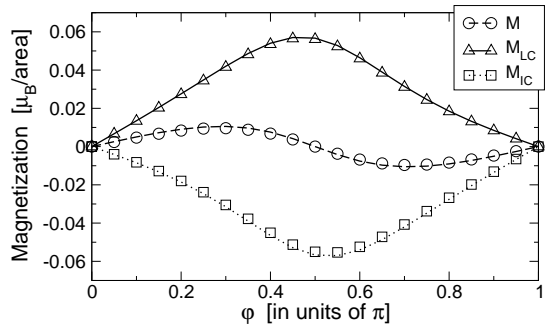}
\caption{\label{fig:results}Numerical results for the magnetization (in
Bohr magnetons per unit area) for the Haldane model.  Symbols denote
results from a finite 30$\times$30 sample, while curves represent
results of a k-space calculation on a periodic system.}
\end{figure}

The results of our numerical calculations for a finite 30$\times$30
sample are depicted as the symbols in Fig.~\ref{fig:results}. First, we
calculated the total magnetic moment according to  Eq.~(\ref{m_tot_1})
and divided by the total sample area $A$ to obtain the magnetization
$M$ indicated as circles in Fig.~\ref{fig:results}.  Next, we evaluated
the contribution to Eq.~(\ref{m_tot_1}) from a single WF deep in the
bulk of the finite sample and divided by the unit-cell area $A_0$ to
obtain a ``local circulation'' magnetization $M\sLC$ plotted as
triangles in Fig.~\ref{fig:results}.  We expected these two quantities
to agree with each other within some numerical tolerance. On the
contrary, the results indicate no agreement whatsoever.

This surprising result forced us to reconsider our entire line of
argument, revealing a profound oversight. It is easily shown that each
bulk band of an insulating crystal must carry no net current, even in
the absence of time-reversal symmetry.  This means that each bulk-like
WF, such as the one that we chose from the deep interior of the sample,
must carry no net current, as is easily confirmed in the numerical
calculation.  We had assumed that the WFs at the boundary of the finite
sample would likewise carry no net current, but \emph{this assumption
is incorrect}.  In fact, the WFs near the boundary do carry a net
current, and the total circulation associated with these net currents
provides just the needed contribution to resolve the discrepancy.

This is illustrated in Fig.~\ref{fig:model}(b), where we have plotted
the net current $-e\v_i=-e\me{\phi_i}{\v}{\phi_i}$ located at the
Wannier center $\rb_i=\me{\phi_i}{\r}{\phi_i}$ for each WF in the
sample. While confirming that the net currents in the deep interior are
exponentially small, the results reveal that the WFs near the surface
do carry a substantial current that contributes extensively to the
total magnetic moment of the sample.  Dividing by the sample area, we
obtain an ``itinerant circulation'' contribution $M\sIC$ to the
magnetization that is plotted as squares in Fig.~\ref{fig:results}. A
glance at the figure suggests, and numerical tests confirm, that
$M=M\sLC+M\sIC$ within numerical precision~\cite{explan-hald-symm}.

It is not surprising that $M\sLC$, corresponding to the local
circulation of a bulk WF, is a bulk property of the
insulator~\cite{ChemPhysChem}. It is far less clear whether $M\sIC$ is
also a bulk property.  To show that it is, we go back to
Eq.~(\ref{m_tot_1}). For simplicity, we restrict ourselves henceforth
to the case of a two dimensional system with a single occupied band
(the generalization to three dimensions is straightforward, while the
multiband treatment is more subtle). Then Eq.~(\ref{m_tot_1}) can be
rewritten as
\beq
M = -\frac{e}{2Ac}\sum_i\Big[
    \underbrace{\me{\phi_i}{(\r-\rb_i)\times\v}{\phi_i}}_{\text{LC}} +
    \underbrace{\rb_i\times\me{\phi_i}{\v}{\phi_i}}_{\text{IC}}\Big]
    \;,\label{m_tot_2}
\eeq
where again `LC' and `IC' correspond to local and itinerant circulation
contributions, respectively.  We divide the finite sample into an
``interior'' and a ``surface'' region in such a way that the latter
occupies a non-extensive fraction of the total sample area in the
thermodynamic limit. We label WFs from the surface and interior regions
as $\ket{\phi_s}$ and $\ket{\R}$, respectively, where $\R$ is a lattice
vector.

Next, we note that the WFs in the surface region make a negligible
contribution to the LC term of Eq.~(\ref{m_tot_2}) since they occupy a
non-extensive fraction of the area in the thermodynamic limit. The LC
term then becomes
\bea
M\sLC &=& -\frac{e}{2Ac}\sum_\R \me{\R}{(\r-\bar{\r}_{\R})\times\v}{\R}\nn
      &=& -\frac{e}{2A_0c} \me{\0}{\r\times\v}{\0}\;.\label{MLC}
\eea
Here we have used translational symmetry and the fact that
$\me{\R}{\v}{\R}$=0 (since bulk bands carry no net current).
Eq.~(\ref{MLC}) shows that $M\sLC$ can be expressed simply in terms of
the bulk WF $\ket{\0}$ in the home unit cell. Turning to the IC term,
the interior WFs now make no contribution (again because
$\me{\R}{\v}{\R}$=0) so that
\beq
M\sIC = -\frac{e}{2Ac}\sum_s \rb_s\times\v_s\;,\label{MICa}
\eeq
where the sum runs over surface WFs only.

\begin{figure}
\includegraphics[width=2.4in]{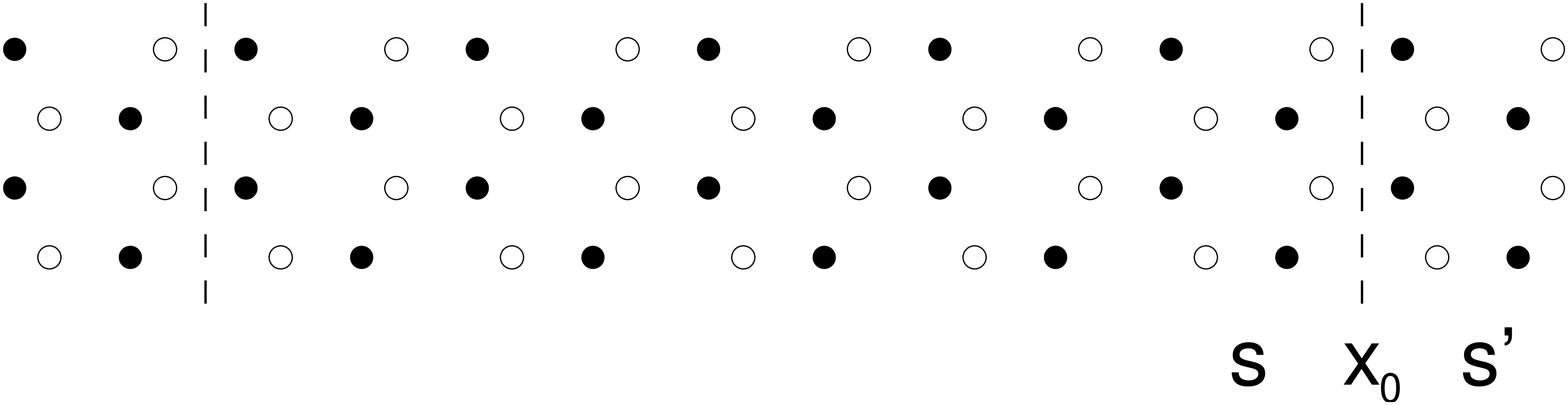}
\caption{\label{fig:strip}Horizontal slice from a sample that extends
indefinitely in the vertical direction but is otherwise similar to the
one in Fig.~\ref{fig:model}.  Vertical dashed lines delimit bulk and
surface regions in which WFs are labeled by $s$ and $s'$,
respectively.}
\end{figure}

We now concentrate on this IC contribution and consider a vertical
strip of which one horizontal section is sketched in
Fig.~\ref{fig:strip}, and choose vertical boundaries (dashed lines) to
discriminate between interior and surface regions.  Focusing on the
right edge, we use labels $s$ and $s'$ to label WFs in the interior
($x<x_0$) and surface ($x>x_0$) regions, respectively. The vertical
macroscopic current flowing in the right surface is 
\beq
I_y = -\frac{e}{\Delta l}{\sum_{s'}}' v_{s',y} \;, \label{piece}
\eeq
where the primed  sum is further limited to WFs whose centers are
inside a vertical segment of length $\Delta l$, equal to the vertical
repeat unit. The current carried by the $i$-th WF can be written in
terms of contributions from its neighbors as
\beq
\v_i = -\frac{i}{\hbar}\,\me{\phi_i}{[\r,H]}{\phi_i}
     = \sum_{j} \v_\bji\;,
\eeq
where $\v_\bji = (2/\hbar) \,{\rm Im}\, \r_{ij}\,H_{ji}$ has the
interpretation of a current ``donated from WF $j$ to WF $i$.'' Since
$\v_\bji=-\v_\bij$, the total current carried by any subset of WFs can
be computed as the sum of all $\v_\bji$ for which $i$ is inside and $j$
is outside the subset.   Applying this to the piece of surface region
considered above, Eq.~(\ref{piece}) becomes 
\beq
I_y = -\frac{e}{\Delta l}\,{\sum_{s}} {\sum_{s'}}' v_{\ev{s,s'},y}\;.
\label{piece2}
\eeq
Setting the boundary deep enough below the surface to be in a bulk-like
region and invoking the exponential localization of the WFs and of
derived quantities like $\v_\bji$, we can identify $\ket{\phi_s}$ and
$\ket{\phi_{s'}}$ with the corresponding bulk WFs. Exploiting
translational symmetry, $\v_{\ev{\R,\R'}} = \v_{\ev{0,\R'-\R}}$,
Eq.~(\ref{piece2}) becomes
\beq
I_y = -\frac{e}{\Delta l}\,{\sum_{R_x < x_0}}\;
      {\sum_{R'_x > x_0}}^{\!\!\!\prime} v_{\ev{0,\R'-\R},y} \;,
      \label{piece3}
\eeq
where the sum is still restricted to a segment of height $\Delta l$. 
The number of terms in Eq.~(\ref{piece3}) having a given value of
$\R'-\R$ is just $(R'_x - R_x) \Delta l/A_0$ if $(R'_x-R_x)>0$ and zero
otherwise. With a change of summation index, Eq.~(\ref{piece3}) becomes
\beq
I_y = -\frac{e}{2A_0}\,{\sum_{\R}} R_x v_{\ev{0,\R},y} \;,
    \label{piece4}
\eeq
where the factor of 2 enters because the sum has been extended to all
$\R$. For a boundary fragment of arbitrary orientation,
Eq.~(\ref{piece4}) generalizes to $I_\alpha = \sum_\beta
G_{\alpha\beta}\,\hat{n}_\beta$, where $\nhat$ is the unit normal to
the boundary, and
\beq
G_{\alpha\beta} = -\frac{e}{A_0\hbar} \sum_\R\,
                  {\rm Im}\,\me{\R}{r_\alpha}{\0}\me{\0}{H}{\R}\,
                  R_\beta\;.\label{Gdef}
\eeq
The contribution of this itinerant current to the magnetic moment
$(1/2c)\oint (\r\times\I) \, dl$ is easily seen to be related to the
antisymmetric part~\cite{explan-symmpart} of $G$, so that
\beq
M\sIC =\frac{1}{c}\, G_{xy}^A=\frac{1}{2c}\,(G_{yx}-G_{xy})\;.\label{MICr}
\eeq
Equations~(\ref{Gdef}) and (\ref{MICr}) constitute our first major
result, showing that the itinerant circulation contribution to the
orbital magnetization can indeed be expressed as a bulk property in
terms of the bulk WFs alone.

In the remainder of this Letter, we show that the two contributions
$M\sLC$ and $M\sIC$ can both be converted into k-space expressions that
can be evaluated directly in the Bloch representation.  The WFs are
defined via
\beq
\ket{\R} = \ltp\int d^2k\,e^{i\k\cdot(\r-\R)}\,
           \ket{u_{\k}}\;,\label{wdef}
\eeq
where $\ket{u_{\k}}=e^{-i\k\cdot\r}\,\ket{\psi_{\k}}$ is the
cell-periodic part of the Bloch function $\ket{\psi_{\k}}$. Inserting
Eq.~(\ref{vdef}) into Eq.~(\ref{MLC}) and using $\r\times\r=0$, it
follows that
\beq
M\sLC = \frac{e}{2A_0\hbar c}\,{\rm Im}\, \me{\0}{\r\times H\r}{\0}\;.
\label{MLC0}
\eeq
Defining $H_\k=e^{-i\k\cdot\r}He^{i\k\cdot\r}$ and using that
\beq
\r\,\ket{\R} = i\,\ltp\int d^2k\,e^{i\k\cdot(\r-\R)}\,
               \ket{\park u_\k}\;,\label{xw}
\eeq
Eq.~(\ref{MLC0}) becomes
\beq
M\sLC = \frac{e}{2\hbar c} \,{\rm Im}  \int\frac{d^2k}{(2\pi)^2}\; 
        \me{\park u_\k}{\times H_\k\,}{\park u_\k}\;,\label{MLC1}
\eeq
in agreement with Ref.~\cite{ChemPhysChem}.

In order to convert $M\sIC$ to k-space, we note that the matrix
elements appearing in Eq.~(\ref{Gdef}) are given by
\bea
\me{\0}{\r}{\R}   &=& \ltp\int d^2k\;
                      \A_\k\,e^{-i\k\cdot\R}\;,\\
\me{\0}{H}{\R}    &=& \ltp\int d^2k\; E_\k\,e^{-i\k\cdot\R}\;,
\eea
where the Berry connection $\A_\k = i\,\me{u_\k}{\park}{u_\k}$ is real
and $E_\k$ is the band energy. After some algebra including an
integration by parts, we find
\beq
G\sab = -\frac{e}{\hbar}\int\ddkk\; E_\k \,\parb A_{\k\alpha}\;.
        \label{Gab2}
\eeq
Inserting in Eq.~(\ref{MICr}) gives
\beq
M\sIC = -\frac{e}{2\hbar c} \int\ddkk\;E_\k\,\Omega_\k\;,
\label{MIC1}
\eeq
where $\Omega_\k=\park\times\A_\k$ is the Berry curvature.

Interestingly, both magnetization contributions (\ref{MLC1}) and
(\ref{MIC1}) are individually gauge-invariant,
i.e., insensitive to the choice of phases of the Bloch functions used to
construct the WFs.  This was shown for Eq.~(\ref{MLC1}) in
Ref.~\cite{ChemPhysChem}, and it follows immediately for
Eq.~(\ref{MIC1}) because $\Omega(\k)$ is a gauge-invariant quantity.
Each contribution is also invariant with respect to a
shift of the zero of the Hamiltonian; the effect of such a shift
is proportional to $\int d^2k\,\Omega_\k=2\pi\,C$, where
$C$ is the Chern number which has been assumed to vanish.

Adding $M\sLC$ and $M\sIC$, placing both in a common form, and
returning to three dimensions, we find that the total magnetization of
the crystalline solid can be expressed as
\beq
\M = \frac{e}{2\hbar c} \,{\rm Im} \int \frac{d^3k}{(2\pi)^3} \,
    \me{\park u_\k}{\times (H_\k+E_\k)\,}{\park u_\k}\;.\label{mtot}
\eeq
This is our principal result. We note that Eq.~(\ref{mtot}) is
consistent with Eq.~(11) of Ref.~\cite{Niu05}, thereby providing a
direct, fully quantum derivation of a result inferred there on the
basis of semiclassical arguments alone.

To confirm the correctness of our formulation, we used
discretized~\cite{Sai02,ChemPhysChem} versions of Eqs.~(\ref{MLC1}) and
(\ref{MIC1}) to calculate $M\sLC$, $M\sIC$, and $M$ for the Haldane
model using a 300$\times$300 k-point mesh. The results, drawn as the
lines in Fig.~\ref{fig:results}, are entirely consistent with the
results for finite samples.  We also carried out numerical tests which
confirmed that the extrapolation of the results from $N\times N$ finite
samples for $N=10$, 20 and 30 to $N\rightarrow\infty$ shows very
precise agreement with the k-space calculation on a dense mesh.  We can
thus be confident that the formal derivations are correct and that
there is no longer any possibility that terms in the magnetization are
being overlooked.  We have carried out similar tests for other
tight-binding models with reduced symmetry~\cite{explan-hald-symm},
with similar results.

In conclusion, we have derived a formula for the orbital magnetization
of a crystalline system by working in the Wannier representation, and
we have demonstrated its correctness via numerical tests.  While
limited to the case of a non-interacting zero-Chern-number insulator in
a vanishing (or commensurate) magnetic field, our result nonetheless
represents significant progress towards a more general theory of
orbital magnetization. The resulting formula is easily evaluated in the
context of a k-space electronic-structure code.  The generalization to
the multiband case will be discussed in a forthcoming publication.  It
remains tantalizingly uncertain whether such a Wannier-based approach
can also be generalized to handle insulators with non-zero Chern
numbers, metals, or arbitrary magnetic fields.

This work was supported by ONR grant N00014-03-1-0570 and NSF grant
DMR-0233925.


\begin{thebibliography}{10}
\bibitem{explan-hald-symm}
   The antisymmetry of the total magnetization $M$ about
   $\varphi$=$\pi/2$ in Fig.~\ref{fig:results} is related to the
   presence of a subtle particle-hole symmetry in the Haldane
   model~\cite{Haldane88,ChemPhysChem}.
\bibitem{explan-symmpart}
   The corresponding symmetric part $G_{xy}^S$ is related to the time
   derivative of the spread of Wannier functions~\cite{Marzari97}.
\end{thebibliography}
\end{document}